\documentstyle[12pt,a4] {article}
\newcommand{\beq}{\begin{equation}}
\newcommand{\eeq}{\end{equation}}
\newcommand{\beqa}{\begin{eqnarray}}
\newcommand{\eeqa}{\end{eqnarray}}
\newcommand{\ba}{\begin{array}}
\newcommand{\ea}{\end{array}}

\begin{document}

\begin{center}
{\large \bf Quantum chaos in A=46--50 atomic nuclei}
\end{center}

\vskip 0.5 truecm

\begin{center}
{\bf E. Caurier$^{(a)}$, {J.M.G. G\'omez}$^{(b)}$,
V.R. Manfredi$^{(c)}$, L. Salasnich$^{(b)}$}

\vskip 0.5 truecm

$^{(a)}$ Groupe de Physique Th\'eorique, Centre de Recherches Nucl\'eaires, \\
IN2P3--CNRS/ Universit\'e Louis Pasteur, \\
BP 20, F--67037 Strasbourg--Cedex, France

\vskip 0.5 truecm

$^{(b)}$ Departamento de F\'\i sica At\'omica, Molecular y Nuclear, \\
Facultad de Ciencias F\'\i sicas, Universidad Complutense de Madrid, \\
E--28040 Madrid, Spain \\
E-mail address: Gomezk@eucmax.sim.ucm.es, Telefax: +34-1-394 51 93
\vskip 0.5 truecm

$^{(c)}$ Dipartimento di Fisica "G. Galilei" dell'Universit\`a di Padova, \\
INFN, Sezione di Padova, \\
Via Marzolo 8, I--35131 Padova, Italy \\
Interdisciplinary Laboratory, International School for Advanced Studies, \\
Strada Costiera 11, I--34014 Trieste, Italy

\end{center}

{\bf Abstract:} The spectral statistics of low--lying states of several
$fp$ shell nuclei are studied with realistic shell--model
calculations.
For Ca isotopes, we find significant deviations from the predictions of the
random--matrix theory which suggest that some spherical nuclei
are not as chaotic in nature as the conventional view assumes.



\newpage

\par
In recent years many authors have shown great interest in the fluctuation
properties of energy levels. It is well known that the fluctuation
properties of quantum systems with underlying classical chaotic behaviour and
time--reversal symmetry agree with the predictions of the Gaussian
Orthogonal Ensemble (GOE) of random matrix theory, whereas
quantum analogs of classically integrable systems display the characteristics
of the Poisson statistics [1].
\par
One of the best systems for the study of
quantum chaos is the atomic nucleus, which was studied
in pioneer work [2] and has been the subject of many investigations [3].
In atomic nuclei, the fluctuation properties of experimental energy levels are
best studied in the domain of neutron and proton resonances near the nucleon
emission threshold, where a large number of levels of the same spin and parity
in the same nucleus are present,
and an excellent agreement with GOE predictions
has been found [2]. In the ground state region, however, the samples of
consecutive experimental levels of the same spin and parity in any one
nucleus are quite small. Therefore it is more difficult to calculate
reliable mean values and fluctuations of statistics such as energy level
spacings. In order to circumvent this difficulty,
in recent years [4--6]
statistical analyses of experimental low energy levels have
combined data from a large range of excitation energies and angular
momenta of a nucleus or a set of nuclei. Such analyses have provided
evidence suggesting that spherical nuclei show level spectra close to GOE
predictions and deformed nuclei show strong deviations from GOE behaviour.
In a recent analysis [7] of level spacings close to the yrast line of
deformed nuclei with $Z=62$--$75$ and $A=155$--$185$,
the average level spacing for
states with the same spin and parity was calculated for the total ensemble
instead of for individual nuclei. The level spacing fluctuations
obtained are quite close to the Poisson distribution, showing evidence of
regular motion, although it is possible that this Poisson behaviour may be
related to a hidden symmetry in the $K$ quantum number.
In general it seems quite clear that the available experimental data are
insufficient to establish the borderlines in mass number, excitation energy,
etc., between order and chaos in nuclei.
\par
Theoretical work, especially realistic
shell model calculations, should help to establish the domains of chaos
in nuclei. The statistical analyses of shell--model energy
spectra and wave functions have been mainly concentrated on the $sd$
shell region, and a very chaotic behaviour has been found for
these nuclei [8--11]. However, recent shell--model
calculations by Bae {\it et al.} [12] have shown that for
nuclei of mass $A=212$ shell--model spectra display
features of regular motion in some cases. The main reason
for this seems to be the relatively small values of the residual
interaction matrix elements as compared to the average spacing between
neighboring single--particle levels in heavy nuclei.
\par
In this work, we undertake the statistical analysis of the shell--model
energy levels in the $A=46$--$50$ region. Exact calculations
are performed in the ($f_{7/2}$,$p_{3/2}$,$f_{5/2}$,$p_{1/2}$)
shell--model space, assuming a $^{40}$Ca inert core [13].
The diagonalizations are performed in the {\it m}--scheme using a fast
implementation of the Lanczos algorithm with the code ANTOINE [14].
For a fixed number of valence protons and neutrons
we calculate the energy spectrum for
projected total angular momentum $J$ and total isospin $T$.
The interaction we use is a minimally modified Kuo--Brown
realistic force with monopole improvements [15].
\par
We calculate the $T=T_z$ states from $J=0$ to $J=9$
for all the combinations of $6$ active nucleons,
i.e. $^{46}$V, $^{46}$Ti, $^{46}$Sc, and $^{46}$Ca, and also
for $^{48}$Ca and $^{50}$Ca.
Table 1 shows the {\it m}--scheme dimensions and the largest dimensions
in the $JT$ scheme of the isotopes analyzed.
\par
Since we are looking for deviations from chaotic features, we are
mainly interested in the low--lying levels, up to a few MeV above the $JT$
yrast line. Let us consider the energy levels up to $4$, $5$ and $6$ MeV above
the yrast line, and calculate the fluctuations around the
average spacing between neighboring levels. In this range of energies,
the level spectrum can be mapped into unfolded levels with
quasi--uniform level density by using the constant temperature
formula. In order to guarantee that the results up to different energies
are unaffected by the unfolding procedure, the unfolding is
performed using always the whole set of levels up to 6 MeV,
for each $JT$ set in the nucleus.
In some cases, the resulting number of levels is too small,
e.g. 3 levels for $J=0$ in $^{46}$Ca, and then we use a larger set of
levels for the calculation of the mean level spacing as a function
of energy. The mean level density can be assumed to be of the form
\beq
{\bar \rho}(E)={1\over T}\exp{[(E-E_0)/T]} ,
\eeq
where $T$ and $E_0$ are constants. For fitting purposes it is better
to use not ${\bar \rho}(E)$ but its integral ${\bar N}(E)$. We write
\beq
{\bar N}(E)=\int_0^E {\bar \rho}(E')dE' + N_0 = \exp{[(E-E_0)/T]}-
\exp{[-E_0/T]}+N_0 .
\eeq
The constant $N_0$ represents the number of levels with energies less
than zero. Following Shriner {\it et al.} [5], we consider Eq. 2 as an
empirical function to fit the data and let $N_0$ take non--zero values.
The parameters $T$, $E_0$ and $N_0$
that best fit $N(E)$ are obtained by minimizing the function:
\beq
G(T,E_0,N_0)=\int_{E_{min}}^{E_{max}} [N(E)-{\bar N}(E)]^2 dE ,
\eeq
where $N(E)$ is the number of levels with energies less than or equal to $E$.
The energies $E_{min}$ and $E_{max}$ are taken as the first and last
energies of the level sequence. As an example, Fig. 1 illustrates
the fit to the integrated level density $N(E)$ for the $J^{\pi}T=6^+1$
levels of $^{46}$Ti.
\par
The spectral statistic $P(s)$ is used
to study the local fluctuations of the energy levels [16,17].
$P(s)$ is the distribution of nearest--neighbour spacings
$s_i={\bar N}(E_{i+1})-{\bar N}(E_i)$ of the unfolded levels.
It is obtained by accumulating the number of spacings that lie within
the bin $(s,s+\Delta s)$ and then normalizing $P(s)$ to unity.
\par
For quantum systems whose classical analogs are integrable,
$P(s)$ is expected to follow the Poisson limit, i.e.
$P(s)=\exp{(-s)}$. On the other hand,
quantal analogs of chaotic systems exhibit the spectral properties of
GOE with $P(s)= (\pi / 2) s \exp{(-{\pi \over 4}s^2)}$ [1,2].
\par
The distribution $P(s)$ is the best spectral statistic to analyze
shorter series of energy levels and the intermediate regions
between order and chaos.
In order to quantify the chaoticity of $P(s)$ in terms of a parameter,
it can be compared for example to the Brody or the Berry--Robnik
distributions, which are adequate for the description of
intermediate situations between order and chaos.
Although each of these distributions have some advantage in
limiting cases [18], they are very similar in a particular case like ours.
We use here the Brody distribution [19], given by
\beq
P(s,\omega)=\alpha (\omega +1) s^{\omega} \exp{(-\alpha s^{\omega+1})},
\eeq
with
\beq
\alpha = (\Gamma [{\omega +2\over \omega+1}])^{\omega +1}.
\eeq
This distribution interpolates between the Poisson distribution ($\omega =0$)
of integrable systems and the GOE distribution ($\omega =1$) of
chaotic ones, and thus the parameter $\omega$ can be used as a simple
quantitative measure of the degree of chaoticity.
\par
In order to obtain more meaningful statistics, $P(s)$ is calculated using the
unfolded level spacings of the whole set of $J=0$--$9$ levels
for fixed $T$ up to a given energy limit above the yrast line.
The number of $J=0$--$9$ spacings below $4$, $5$ and $6$ MeV range
from $42$, $66$ and $105$ in $^{46}$Ca, to $86$, $149$ and $231$ in
$^{46}$Ti, respectively.
\par
Table 2 shows the best fit Brody parameter $\omega$ of the $P(s)$
distribution for the $J=0$--$9$ set of
level spacings in the $A=46$ nuclei up to $4$, $5$ and $6$ MeV above
the yrast line. Clearly, $^{46}$V, $^{46}$Ti and $^{46}$Sc
are chaotic for these low energy levels,
but there is a considerable deviation from GOE
predictions in $^{46}$Ca, which is a single closed--shell nucleus.
In view of the peculiarity of this nucleus, we performed
calculations for $^{48}$Ca and $^{50}$Ca, and obtained again strong deviations
toward regularity, as the values of Table 2 show.
Thus, for the Ca isotopes we find the same kind of phenomenon
obtained by Bae {\it et al.} [12]
in the heavy single--closed nuclei $^{212}$Rn and $^{212}$Pb,
namely that low--lying states deviate strongly from chaoticity
toward regularity.
\par
To obtain a better estimate of the Brody parameter, we can combine
spacings of different nuclei. In Fig. 2 (a) we plot $P(s)$ for
$^{46}$V+$^{46}$Ti+$^{46}$Sc, and in Fig. 2 (b) for
$^{46}$Ca+$^{48}$Ca+$^{50}$Ca, up to $6$ MeV above the yrast lines.
The number of level spacings is now sufficiently large to yield
meaningful statistics and we see that Ca isotopes are not very
chaotic at low energy, in contrast to other nuclei in the same region.
\par
Concerning the energy dependence of $\omega$ in the low energy region,
for individual nuclei we see fluctuations which may be due to the fact
that some of the samples of levels are not very large.
Table 2 shows the results for $^{46}$V+$^{46}$Ti+$^{46}$Sc and
$^{46}$Ca+$^{48}$Ca+$^{50}$Ca. For these larger sets of levels,
fluctuations are reduced, and the energy dependence of $\omega$ is seen
to be small in the $4$--$6$ MeV range, although the chaoticity
increases slightly with the excitation energy.
\par
Since the semi--closed nuclei are quite regular for the low--lying levels,
compared with other neighboring nuclei, it is interesting to
find out in the framework of the model the chaoticity at higher
excitation energies. Therefore we have calculated the whole spectrum for
$^{46}$Ca, unfolding the levels with the local density method [3].
Different $J$ values were combined to improve the $P(s)$ statistics.
The value of $\omega$ increases rather smoothly with the excitation energy
end point, and we obtain $\omega =0.72$, $0.83$ and $0.88$ for levels
below $12$, $16$ and $20$ MeV, respectively.
We do not find significant differences in $\omega$
for different $J$ values in $^{46}$Ca when they are analyzed separately.
\par
Why are Ca isotopes less chaotic than their neighbors?
We observe that the two--body matrix elements of the
proton--neutron interaction are, on average, larger than those of
the proton--proton and neutron--neutron interactions.
Consequently the single--particle mean--field motion in nuclei with both
protons and neutrons in the valence orbits suffers
more disturbance and is thus more chaotic.
\par
It should be noted that an analysis [20] of experimental energy levels
below $4.3$ MeV excitation energy in the semi--magic nucleus
$^{116}$Sn yields a near--neighbor spacing distribution which is
intermediate between GOE and Poisson, with
$\omega =0.51 \pm 0.19$. This result is consistent with the theoretical
findings of Bae et al. [12] for $^{212}$Rn and $^{212}$Pb, and
our present results for Ca isotopes.
\par
The energy spectra of all these nuclei are the result of the interaction
of like nucleons, and we conclude that this residual interaction is
relatively weak and, at least for low energy levels, it
only partially destroys the regular motion of nucleons generated
by the nuclear mean field.
\par
We can ask ourselves why do all shell--model
calculations give chaotic features for $sd$ shell
nuclei, without any significant deviations towards regularity.
First, it
should be noted that most of these calculations include a large
number of states, up to excitation energies far above the nucleon
emission threshold. This is, for example, the case of the $^{22}$O
calculations of Bae {\it et al.} [12], which include the full set of
levels for several $J$ values and obtain $\omega =0.96$.
Second, we notice that Ormand and Broglia [10]
obtained a GOE--like distribution
for the first two spacings of each spectrum for a set of $sd$ shell
nuclei. However,
these nuclei have valence protons and neutrons and are
thus similar to the case of $^{46}$V, $^{46}$Ti and $^{46}$Sc,
for which we also find chaotic behaviour.
We conclude that no regular features have been found
in the $sd$ shell region because single--closed nuclei have a very small
configuration space for these nuclei and thus too few low--lying
levels for statistical analysis,
and also because the disturbance of single--particle motion
by the two--body interaction is greatest in light nuclei, where the range
of the single--particle orbits is not much larger
that the range of the nuclear force.

\vskip 0.5 truecm

\par
This work has been partially supported by MURST and INFN, Italy, and
by DGICYT grant PB93--0263 and CICYT grant AEN--0776, Spain.

\newpage

\parindent=0. pt
\section*{References}

\vskip 0.5 truecm

[1] M. C. Gutzwiller, {\it Chaos in Classical and Quantum Mechanics}
(Springer--Verlag, Berlin, 1990);
A. M. Ozorio de Almeida, {\it Hamiltonian Systems: Chaos and
Quantization} (Cambridge University Press, Cambridge, 1990);
K. Nakamura, {\it Quantum Chaos} (Cambridge Nonlinear Science Series,
Cambridge, 1993)

[2] R. U. Haq, A. Pandey and O. Bohigas, Phys. Rev. Lett. {\bf 48}
(1982) 1086.

[3] O. Bohigas, M. J. Giannoni and C. Schmit, in {\it Quantum Chaos and
Statistical Nuclear Physics}, Ed. T. H. Seligman and H. Nishioka
(Springer--Verlag, Berlin, 1986);
O. Bohigas and H. A. Weidenm\"uller, Ann. Rev. Nucl. Part. Sci.
{\bf 38}, 421 (1988); M. T. Lopez--Arias, V. R. Manfredi,
and L. Salasnich, Riv. Nuovo Cim. {\bf 17}, N. 5, (1994) 1.

[4] G. E. Mitchell, E. G. Bilpuch, P. M. Endt
and J. F. Jr. Shriner, Phys. Rev. Lett. {\bf 61} (1988) 1473.

[5] J. F. Jr. Shriner, E. G. Bilpuch, P. M. Endt and
G. E. Mitchell, Z. Phys. A {\bf 335} (1990) 393.

[6] J. F. Jr. Shriner, G. E. Mitchell and T. von Egidy,
Z. Phys. {\bf 338} (1991) 309.

[7] J. D. Garrett, J. R. German, L. Courtney and J. M. Espino, in
{\it Future Directions in Nuclear Physics}, p. 345,
Ed. J. Dudek and B. Haas (American Institute of Physics,
New York, 1992)

[8] T. A. Brody {\it et al.}, Rev. Mod. Phys. {\bf 53} (1981) 385.

[9] H. Dias, M. S. Hussein, N. A. de Oliveira and B. H. Wildenthal,
J. Phys. G {\bf 15} (1989) L79;
V. Paar {\it et al.}, Phys. Lett. B {\bf 271} (1991) 1.

[10] W. E. Ormand and R.A. Broglia, Phys. Rev. C {\bf 46} (1992) 1710.

[11] V. Zelevinsky, M. Horoi and B. A. Brown, Phys. Lett. {\bf 350} (1995) 141;
M. Horoi, V. Zelevinsky, B. A. Brown, Phys. Rev. Lett. {\bf 74} (1995) 5194.

[12] M. S. Bae, T. Otshka, T. Mizusuki and N. Fukumishi, Phys. Rev. Lett.
{\bf 69} (1992) 2349.

[13] R. D. Lawson, {\it Theory of the Nuclear Shell Model}
(Clarendon, Oxford 1980).

[14] E. Caurier, computer code ANTOINE, C.R.N., Strasbourg (1989);
E. Caurier, A. P. Zuker and A. Poves: in {\it Nuclear Structure of
Light Nuclei far from Stability: Experiment ad Theory}, Proceedings of
the Obernai Workshop 1989, Ed. G. Klotz (C.R.N, Strasbourg, 1989).

[15] E. Caurier, A. P. Zuker, A. Poves and G. Mart\'\i nez--Pinedo,
Phys. Rev. C {\bf 50} (1994) 225.

[16] F. J. Dyson and M. L. Mehta, J. Math. Phys. {\bf 4} (1963) 701.

[17] O. Bohigas and M. J. Giannoni, Ann. Phys. (N.Y.) {\bf 89} (1975) 393.

[18] E. Caurier, B. Grammaticos and A. Ramani, J. Phys. A {\bf 23}
(1990) 4903; T. Prosen and M. Robnik: J. Phys. A {\bf 27} (1994) 8059.

[19] T. A. Brody, Lett. Nuovo Cimento {\bf 7} (1973) 482.

[20] S. Raman {\it et al.}, Phys. Rev. C {\bf 43} (1991) 521.

\newpage

\begin{center}
{\bf Table 1}
\end{center}
\vskip 0.5 truecm

Matrix dimensions in the {\it m} scheme and maximum dimensions
in the $JT$ scheme for the analyzed isotopes.

\vskip 1. truecm

\begin{center}
\begin{tabular}{|ccccccc|} \hline\hline
{}~~ & $^{46}$V & $^{46}$Ti & $^{46}$Sc & $^{46}$Ca & $^{48}$Ca & $^{50}$Ca
\\ \hline
{\it m}--scheme dimension
& $121\; 440$ & $86\; 810$ & $30\; 042$ & $3\; 952$ & $12\; 002$ &
$17\; 276$
\\
largest dim. $J^+T$ & $4^+0$ & $4^+1$ & $4^+2$ & $4^+3$ & $4^+4$ & $4^+5$
\\
$J^+T$ dimension & $4\; 750$ & $8\; 026$ & $3\; 783$ & $615$ & $1\; 755$
& $2\; 468$
\\
\hline\hline
\end{tabular}
\end{center}

\newpage

\begin{center}
{\bf Table 2}
\end{center}
\vskip 0.5 truecm

Brody parameter $\omega$ for the nearest neighbour level spacings distribution
for $0\leq J\leq 9$, $T=T_z$ states up to $4$, $5$ and $6$ MeV above the
yrast line in the analyzed nuclei.

\vskip 1. truecm

\begin{center}
\begin{tabular}{|ccccccccc|} \hline\hline
Energy & $^{46}$V & $^{46}$Ti & $^{46}$Sc & $^{46}$Ca & $^{48}$Ca &
$^{50}$Ca & $^{46}$V+$^{46}$Ti+$^{46}$Sc & $^{46}$Ca+$^{48}$Ca+$^{50}$Ca \\
\hline
$\leq 4$ MeV & 1.14 & 0.90 & 0.81 & 0.41 & 0.58 & 0.67 & 0.92 & 0.56\\
$\leq 5$ MeV & 1.10 & 0.81 & 0.96 & 0.53 & 0.58 & 0.69 & 0.93 & 0.60\\
$\leq 6$ MeV & 0.93 & 0.94 & 0.99 & 0.51 & 0.66 & 0.62 & 0.95 & 0.61\\
\hline\hline
\end{tabular}
\end{center}

\newpage

\parindent=0. pt
\section*{Figure Captions}

\vskip 0.5 truecm

{\bf Figure 1}: Best fit of the constant temperature formula
to the integrated level density for $J^{\pi}T=6^+1$ of $^{46}$Ti.

{\bf Figure 2}: $P(s)$ distribution for low--lying
levels of $fp$ shell nuclei with
$0\leq J \leq 9$: (a) $^{46}$V, $^{46}$Ti and $^{46}$Sc;
(b) $^{46}$Ca, $^{48}$Ca and $^{50}$Ca.
The dotted, dashed and solid curves stand for GOE,
Poisson, and Brody distributions, respectively.

\end{document}